\documentclass{nature}
\usepackage{graphicx}
\makeatletter
\let\saved@includegraphics\includegraphics
\AtBeginDocument{\let\includegraphics\saved@includegraphics}
\renewenvironment*{figure}{\@float{figure}}{\end@float}
\makeatother
\usepackage{epstopdf}
\usepackage{dcolumn}
\usepackage{bm}
\usepackage{amssymb}
\usepackage{color}
\usepackage{amsmath}
\usepackage{soul}
\usepackage{float}
\usepackage[normalem]{ulem}

\usepackage{cancel}
\usepackage[dvipsnames]{xcolor}
\usepackage{ulem}

\usepackage[%unicode=true,pdfusetitle,
colorlinks=true]
{hyperref}

\hypersetup{colorlinks,
citecolor=blue,
linkcolor=blue,
urlcolor=blue}

\usepackage[percent]{overpic}
\usepackage[font=scriptsize]{caption}
\title{Generation of Stable Peak-Power Similaritons through Gain-Managed Nonlinearity}

\author{Haoyuan He$^{1,2}$, Maksim Kozlov$^{3}$, Alexander Gabriel Loehr$^{2}$,  Peng Gao$^{4}$, Misha  Ivanov$^{2,5}$,
Pavel Sidorenko$^{5}$}
\begin{document}
\maketitle

\begin{affiliations}
\item Max-Planck-Institut für Kernphysik, Saupfercheckweg 1, 69117 Heidelberg, Germany
\item Max-Born-Institut, Max-Born-Str. 2A, 12489 Berlin, Germany
\item Center for Preparatory Studies, Nazarbayev University, 010000, Astana, Kazakhstan
\item School of Physics, Xidian University, Xi’an 710071, China
\item Technion – Israel Institute of Technology, 3200003 Haifa, Israel
\end{affiliations}

\begin{abstract}
Fiber lasers and amplifiers offer attractive alternatives to conventional solid-state systems. However, generation of high-energy ultrashort laser pulses in fibers faces challenges due to the complex interplay of multiple nonlinear effects arising due to pulse confinement within a small fiber core and also limitations imposed by the gain bandwidth of the available active fibers.  The discovery of  self-similar amplification and gain-managed nonlinear amplification (GMNA)  pulse propagation regimes in fibers with normal  dispersion suggests that these challenges can be turned into an advantage. Here, we show that the GMNA regime can be interpreted as a finite-gain-bandwidth deformation of an ideal stable-peak-power self similar solution.  Our analytical and numerical results show how one should shape the fiber gain as a function 
of propagation length to achieve constant peak power similariton-like pulses with steadily increasing energy, the pulse bandwidth exceeding the gain bandwidth, and the nearly linear frequency chirp allowing for 
efficient pulse compression to its Fourier limit. In the idealized reduced model, neglecting Raman scattering, the SPPS scaling permits ${\mu}J$-level pulse energies. In a realistic Yb-fiber model including Raman scattering, a 60-nJ GMNA pulse is generated and subsequently amplified in a CPA stage to 1 ${\mu}J$ while retaining a spectrum with a transform-limited duration of approximately 45 fs. Our results have significant implications for the fundamental understanding of nonlinear wave dynamics and for the advancement of fiber laser technology, supporting the reliable generation of high-energy pulses for practical use in areas such as micromachining, metrology, and bioimaging. 
\end{abstract}
\maketitle

%\section{Main Text}
Ultrafast fiber lasers and amplifiers have a major impact on science and technology. Although fiber lasers provide advantages such as compactness, good thermo-optical properties, and excellent beam quality, they still lag behind solid-state systems in achievable pulse energy and duration. This limitation arises because confining high-peak-power pulses within a small fiber core over long distances leads to pronounced nonlinear effects that degrade pulse quality. Excessive nonlinear phase shifts can be avoided
by increasing the fiber area to reduce intensity, enabling the highest femtosecond-pulse energies and peak powers obtained with linear chirped-pulse amplification in large-mode area fibers.  While mode-area scaling together with linear chirped-pulse amplification can provide high pulse energies \cite{eidam2010fiber}, The shortest duration attainable through purely linear amplification is constrained by the amplified spectrum and, consequently, by the usable gain bandwidth of the active fiber. Direct nonlinear amplification with broadband seeding has
also reached multi-microjoule energies, including 2.7-$\mu$J pulses
compressed to 76~fs \cite{guo2025monolithic}.

    One route to making further progress is to use 
nonlinearity rather than avoid it. Ideally, one would
want to use the nonlinearity to smoothly broaden the pulse spectrum without introducing strong modulations into the pulse spectral or temporal profile, while exploiting the orders-of-magnitude enhancement of the intensity-length product in a waveguide compared to a freely propagating  Gaussian beam. This underscores the importance of the discovery of 
the so-called self-similar pulse amplification termed "similariton" \cite{kruglov2000self,fermann2000self,kruglov2002self,kruglov2003exact,kruglov2005exact,kruglov2006asymptotically,dudley2007self,oktem2010soliton}. A similariton emerges under specific conditions involving nonlinearity, normal dispersion, and ideally an unbounded gain spectrum. As it propagates, its spectral bandwidth broadens while preserving a linear frequency chirp, and it avoids wave breaking due to its parabolic intensity profile. 

In practice, the maximum spectral width of the  similariton  is constrained by the gain bandwidth \cite{fu2017limits}.
However, in 2019 Sidorenko et al. discovered that it is possible to extend nonlinear pulse evolution beyond the limits imposed by the gain bandwidth \cite{sidorenko2019nonlinear,sidorenko2020generation}
while maintaining the smooth spectral profile characteristic of similariton-like pulses. With long and highly-doped fibers, a seed pulse with central wavelength near the transition between the absorption and gain will first undergo self-phase modulation (SPM) dominated propagation, with the spectral bandwidth growing within the gain bandwidth.  Next, instead of developing the highly structured shape typical of the SPM-dominated regime, the pulse evolves into a smooth temporal profile with an almost linear chirp. At the same time, its spectrum continues to broaden (primarily toward longer wavelengths), eventually exceeding the gain bandwidth. As a result, the pulse can be compressed close to its transform limit, reaching durations below 40 fs using standard diffraction gratings. This allows peak powers an order of magnitude higher than those achievable in conventional self-similar amplifiers. 

Numerical simulations that account for the self-consistent evolution of the gain, calculated using rate equations together with pulse propagation, show that the gain spectrum and the pulse continuously reshape one another as the pulse travels through the fiber. This regime, known as gain managed nonlinear amplification (GMNA), differs from other amplification regimes in that the peak power initially increases, then stabilizes, and gradually decreases during propagation, while the pulse energy continues to grow. More recent experiments have extended nonlinear gain-managed amplification to compact all-fiber architectures and a substantially broader range of pulse energies and repetition rates \cite{xia2025all,zhang2025generation,tang2026compact}.

Recent theoretical studies have provided complementary pictures of the nonlinear dynamics underlying gain-managed amplification \cite{kozlov2025generalized,turitsyn2023nonlinear}. More recently, Wang, Chen, and Wise systematically investigated nonlinear fiber amplification as a function of seed-pulse duration and realistic gain dynamics and placed GMNA within a broader family of nonlinear amplification regimes \cite{wang2026partial}. They showed that gain management is not unique to GMNA: depending on the seed conditions and the spectral region of the Yb gain, it can modify self-similar and partial-parabolic amplification, leading to gain-managed self-similar amplification (GMSSA) and gain-managed partial-parabolic amplification (GMPPA), in addition to GMNA. Their results therefore emphasize that gain management itself does not define a unique propagation regime; rather, the resulting pulse evolution depends on the underlying balance of dispersion, nonlinearity, and the longitudinal and spectral evolution of the gain.

The present work addresses a different but complementary question. Rather than searching for the nonlinear amplification regime that maximizes the compressed peak power for a given seed pulse, we seek a simple analytical propagation law that determines how the longitudinal gain must evolve to compensate dispersive temporal stretching while the pulse energy continues to increase. This analysis leads to an idealized stable peak-power similariton, for which the pulse peak power remains approximately constant during propagation. The SPPS therefore provides a useful reference state for interpreting the evolution of strongly chirped pulses under longitudinal gain management. We subsequently show how the inclusion of a finite, frequency-dependent gain transforms this idealized behavior toward the characteristic asymmetric evolution of GMNA.

Here we develop and substantiate, numerically and analytically, a complementary physical picture of the GMNA regime based on its connection to the SPPS and show why a smooth spectral profile with linear frequency chirp emerges naturally in this regime.  Next, we identify a type of similariton, termed the stable peak-power similariton, which can be achieved through careful gain management as a function of fiber length. This regime enables the generation of high-energy, linearly chirped pulses at the output. Unlike conventional similaritons, the pulse peak power remains constant over long propagation distances while the energy grows steadily. We demonstrate that the GMNA regime can be viewed as a stable peak power similariton modified by the spectral gain profile. Our results show that, absent
Raman nonlinearities, pulses with $\mu J$ scale energies compressible down to about 50 fs can be produced in standard single-mode fibers. As an illustrative application, we examine subsequent chirped-pulse amplification of a simulated GMNA output. The calculated pulse energy reaches 1 $\mu$J while retaining a spectrum whose transform-limited intensity duration is approximately 45 fs.

In fibers with normal dispersion, the similariton is an asymptotic solution of the nonlinear Schrödinger equation (NLSE) with a frequency-dependent but distance-independent gain,
\begin{eqnarray}
%&&
i\frac{\partial}{\partial z}\Psi(z,\tau)
&=&
\frac{\beta_2}{2}\frac{\partial^2}{\partial \tau^2}\Psi(z,\tau)
+i\frac{1}{2}\mathcal{F}^{-1}
\left[
g(\omega)\tilde{\Psi}(z,\omega)
\right]
-\gamma |\Psi(z,\tau)|^2\Psi(z,\tau),
\label{eq:NLSE_1}
\end{eqnarray}
where $\Psi(z,\tau)$ is the slowly varying envelope associated with the optical pulse, $\tilde{\Psi}(z,\omega)=\mathcal{F}{\Psi(z,\tau)}$ is its frequency-domain representation, $g(\omega)$ describes the frequency-dependent gain, $\beta_2$ governs the group-velocity dispersion (GVD), and $\gamma$ is the Kerr nonlinear parameter.

For constant gain, $g_\omega=g$, the similariton naturally develops parabolic intensity profile and a growing spectral bandwidth with quadratic temporal phase, ideal for compression. The similariton drawbacks are (i) its exponentially growing peak power, which at some point cannot be sustained  by the fiber, (ii) limited spectral bandwidth  of the gain and (iii) gain dependence on the propagation length. While gain saturation can be included \cite{kruglov2012parabolic}, finding a route to simultaneously (i) maintain smooth frequency chirp of the pulse, (ii) increase  its spectral bandwidth  beyond the gain bandwidth, and (iii) increase the pulse energy  while maintaining low peak power for realistic
gain conditions remains an outstanding  challenge. The analysis below suggests a route for doing just that.

We take advantage of the strong frequency chirp characteristic for pulses in the GMNA regime\cite{sidorenko2019nonlinear,sidorenko2020generation,kozlov2025generalized}. 
%The chirp links the
%time $\tau$ inside the pulse to its instantaneous frequency $\omega(\tau)$. 
%and connects the pulse representations in time $\psi(\tau)$ and frequency-domain. 
Transforming the Kerr term in 
Eq.(\ref{eq:NLSE_1}) to the frequency domain leads to a convolution, but for a strongly chirped pulse, a straightforward application of the stationary phase method shows that the convolution can be approximated as 
\begin{eqnarray}
\hat{FT}\left[|\psi(z,\tau)|^2 \psi(z,\tau)\right]
\simeq 
|\psi(z,\tau(\omega))|^2\tilde\psi(z,\omega)\simeq 
\frac{1}{\kappa(z)}|\tilde \psi(z,\omega)|^2\tilde\psi(z,\omega)
\label{eq:KerrToOmega}	
\end{eqnarray}
where $\kappa(z)=\partial^2_{\omega,\omega}\tilde\phi(\omega,z)$ is the spectral chirp of the pulse $\tilde\psi(z,\omega)=\tilde A(z,\omega)\exp[i\tilde \phi(z,\omega)]$, and the connection 
between time and frequency is $\tau(z,\omega)=\kappa(z)\omega$. The approximation is valid when the pulse is sufficiently strongly
chirped that the temporal and spectral coordinates are locally related
by stationary phase, i.e. when
$|\kappa(z)|(\Delta\omega)^2 \gg 1$, and when higher-order spectral
phase varies weakly over the stationary-phase bandwidth.

Suppose that by some distance $z=z_0$ the pulse acquires a strong chirp $\kappa_0\equiv\kappa(z_0)=\partial^2_{\omega,\omega}\phi(\omega,z_0)$. Then for $z>z_0$ we can use Eq.(\ref{eq:KerrToOmega}) to approximately transform the NLSE to the frequency domain 
% \begin{eqnarray}
% %&&
% i\frac{\partial}{\partial z}\psi(z,\tau)=
% \frac{1}{2}\beta_2\frac{\partial^2}{\partial \tau^2}\psi(z,\tau)
% %\nonumber \\
% %&& 2\bar{n}_2|A_0|^2\frac{\w_0}{c}
% +i\frac{1}{2} \hat g \psi(z,\tau)
% -\gamma |\psi(z,\tau)|^2\psi(z,\tau)
% \label{eq:NLSE_1a}	
% \end{eqnarray}
\begin{eqnarray}
%&&
i\frac{\partial}{\partial z}\tilde\psi^{(0)}(z,\omega)\simeq
-\frac{\omega^2}{2}\beta_2\tilde\psi^{(0)}(z,\omega)
%\nonumber \\
%&& 2\bar{n}_2|A_0|^2\frac{\w_0}{c}
+i \frac{1}{2} g(\omega,z) \tilde \psi^{(0)}(z,\omega)
-\frac{\gamma}{\kappa(z)}
%\frac{\gamma}{|\phi''_{\omega,\omega}(\omega,z)|}
|\tilde\psi^{(0)}(z,\omega)|^2\tilde\psi^{(0)}(z,\omega)
\label{eq:NLSE_2}	
\end{eqnarray}
where $\tilde\psi^{(0)}(z,\omega)=\tilde A^{(0)}(z,\omega)\exp(i\tilde\phi^{(0)}(z,\omega))$  
is the zero-order approximation 
to the full solution. The solution of  Eq.(\ref{eq:NLSE_2}) is
\begin{eqnarray}
    \tilde\psi^{(0)}(z,\omega) = \tilde A(z_0,\omega) e^{\frac{1}{2}G(\omega,z)} e^{i \bigg\{ \frac{\omega^2}{2}[\kappa_0+\beta_2(z-z_0)] + \gamma |\tilde A(z_0,\omega)|^2 \int_{z_0}^z \frac{\exp(G(\omega,z'))}{(\kappa_0+\beta_2 (z'-z_0))} d z' \bigg\} }
    \label{eq:NLSE_ZeroSolution}
\end{eqnarray}
where
\begin{eqnarray}
    G(\omega,z)=\int_{z_0}^z g(\omega,\tilde{z}) d\tilde{z}
    \label{eq:IntegratedGain}
\end{eqnarray}
is the integrated gain, and we approximated
$\kappa(z)\simeq\kappa_0+\beta_2(z-z_0)$ in the last term of the phase. Equation (\ref{eq:NLSE_ZeroSolution}) retains a Kerr-induced nonlinear phase, however, because the stationary-phase approximation makes the nonlinear term local in frequency, it does not redistribute spectral amplitude and therefore does not reproduce SPM-induced spectral broadening. The resulting $\tilde\psi^{(0)}(z,\omega)$ can then be
Fourier transformed back to the time domain, $\psi^{(0)}(z,\tau)$,
and substituted into the linearized version of
Eq.~(\ref{eq:NLSE_1}),
\begin{eqnarray}
%&&
i\frac{\partial}{\partial z}\psi(z,\tau)\simeq 
\frac{1}{2}\beta_2\frac{\partial^2}{\partial \tau^2}\psi(z,\tau)
%\nonumber \\
%&& 2\bar{n}_2|A_0|^2\frac{\w_0}{c}
+\frac{i}{2}\hat g \psi(z,\tau)
-\gamma |\psi^{(0)}(z,\tau)|^2\psi(z,\tau)
\label{eq:GMA}	
\end{eqnarray}
where it drives the self-phase modulation.
Using Eq.(\ref{eq:GMA}) and Eq.(\ref{eq:KerrToOmega})
to approximate the Kerr term as
$|\psi(z,\tau)|^2\simeq \tilde A^2(z,\omega(\tau))/\kappa(z)$,
we can write the SPM term $\exp{[i\varphi_{SPM}(z,\tau)]}$:
\begin{eqnarray}
%&&
\varphi_{SPM}(z,\tau)\simeq \gamma 
\int_{z_0}^z dz'\frac{1}{\kappa(z')}
|\tilde A^{(0)}(z',\omega(\tau,z')|^2 
=
\gamma 
\int_{z_0}^z dz'\tilde A^2(z_0,\omega(\tau,z'))\frac{e^{G(\omega,z')}}{\kappa(z')} \ \ ,
\label{eq:SPM}	
\end{eqnarray}
where 
$\omega(\tau,z)=\tau/\kappa(z)\simeq \tau/[\kappa_0+\beta_2(z-z_0)]$. In contrast to standard
SPM, the above expression 
incorporates the interplay of dispersion and gain Eq.(\ref{eq:NLSE_ZeroSolution}).
Fig. 1 shows excellent agreement 
between the results of the approximation Eqs.(\ref{eq:NLSE_ZeroSolution},\ref{eq:GMA}) and the direct numerical simulations of the NLSE Eq.(\ref{eq:NLSE_1}) for the typical fiber and input power parameters used in  the GMNA regime.

Eq.(\ref{eq:GMA}) and 
Eqs.(\ref{eq:NLSE_ZeroSolution},\ref{eq:SPM})  allow one to  gauge the interplay of gain, dispersion, SPM, and optimize routes to keeping the peak power limited while increasing the spectral bandwidth. 
For a strongly chirped pulse, the stationary-phase mapping gives
\begin{equation}
P_{\mathrm{pk}}(z) \sim \frac{\tilde{A}^2_{\mathrm{pk}}(z)}{\kappa(z)}.
\end{equation}
Within the zeroth-order approximation of Eq.(\ref{eq:NLSE_ZeroSolution}),
\begin{equation}
\tilde{A}^2_{\mathrm{pk}}(z) \propto \exp[G(z)].
\end{equation}
Requiring the peak power to remain approximately constant therefore gives
\begin{equation}
g(z)=\frac{dG}{dz}\simeq \frac{d}{dz}\left[\ln \kappa(z)\right].
\end{equation}
Using
\begin{equation}
\kappa(z)\simeq \kappa_0+\beta_2(z-z_0),
\end{equation}
we obtain
\begin{equation}
g(z)\simeq
\frac{\beta_2}{\kappa_0+\beta_2(z-z_0)}.
\end{equation}
Thus, asymptotically, the required local gain scales as $1/z$, while the integrated gain scales logarithmically with propagation distance.
% Consequently, the Kerr-related spectral phase in $\tilde\psi^{(0)}(z,\omega)$ Eq.(\ref{eq:NLSE_ZeroSolution}) grows approximately
% linearly with distance, similar to the spectral phase  associated with dispersion. Near the maximum of the pulse spectrum, where the 
% second derivative of $|\tilde\psi(z_0,\omega)|^2$ with respect to $\omega$ is negative, the 
% Kerr term contribution to the spectral chirp compensates 
% dispersion. 
The logarithmic scaling should be regarded as a zeroth-order design rule. In the complete nonlinear propagation, SPM redistributes the spectral intensity, so the optimum gain profile need not satisfy $g_0 z_s = 1$ exactly. Indeed, the parameters used in Fig. 1 correspond to $g_0 z_s = 1.3$, the nearly constant peak power in the full simulation therefore reflects corrections beyond the zeroth-order scaling.

Eq.(\ref{eq:SPM}) shows the role of dispersion 
in generating  the parabolic spectral amplitude and phase
in the GMNA regime. Given that
$\exp{[G(\omega,z)]}/\kappa(z)$ saturates,
the self-phase modulation term becomes determined by the 
spectral amplitude $\tilde A^2(z_0,\omega(\tau,z))$. 
For fixed
$\tau$ and growing $z$, $\omega(\tau,z)=\tau/\kappa(z)\rightarrow 0$.
Hence, the SPM term
necessarily becomes determined by the central part $\omega\rightarrow 0$ of the 
spectral amplitude $A^2(z_0,\omega(\tau,z))$. For an 
initially smooth spectral amplitude, its central part 
by definition has the shape of an inverted parabola, naturally resulting
in the quadratic spectral phase and the inverted parabolic shape of the pulse characteristic for the Kruglov similariton, making the latter a natural attractor even
in media with spectrally limited gain. This explains
the origin of the GMNA behavior.

With this insight, we can identify the regime which we refer to as the "stable peak power similariton" (SPPS). It provides a sustained balance between gain, dispersion, and nonlinearity. In contrast to the traditional similariton with constant gain, the SPPS is described by a prescribed longitudinal gain profile. The idealized model does not require temporal modulation of the pump. In a practical amplifier, longitudinal gain variation can arise through pump depletion and signal-induced changes in population inversion. Whether the required peak-power balance is realized must be determined from the coupled gain–propagation dynamics, rather than assumed from the prescribed profile.
 
The results of the illustrative simulations
with a predefined $\omega,z$-dependent gain profile are shown in Figs. 1,2. They are later  confirmed by the full simulation which includes the self-consistent feedback between the propagating
pulse and the fiber gain (the GMNA regime, see Fig.3). We emphasize that the SPPS should not be interpreted as a general designation for gain-managed nonlinear amplification. As recently demonstrated by Wang \cite{wang2026partial}, longitudinal and spectral gain evolution can modify several distinct nonlinear propagation regimes, including conventional self-similar and partial-parabolic amplification. Here, SPPS refers specifically to the idealized propagation state defined by approximately stationary peak power, increasing pulse energy, and a strongly chirped, near-parabolic pulse under a suitably chosen longitudinal gain profile. The spectrally flat case considered below therefore serves as a reference solution rather than as a complete model of GMNA.

Fig.1 shows the simplest case 
of the frequency-independent, but z-dependent gain. 
The results are obtained by propagating a 0.35-ps (FWHM) Gaussian pulse with an energy of 35 pJ through an 8-meter fiber amplifier. The fiber parameters $\gamma=6\times10^{-3}W^{-1}m^{-1}$ and $\beta_2=35\times10^{-3}ps^2m^{-1}$ reflect the realistic ranges expected in high-gain rare-earth-doped fiber amplifiers.  Fully numerical (Eq.(\ref{eq:NLSE_1}), solid curves) and approximate 
semi-analytical results (Eqs.(\ref{eq:NLSE_ZeroSolution}-\ref{eq:GMA}), rhombuses and squares)
are in excellent agreement. 

The second is the SPPS stage. The key to this stage lies in controlling the gain. We prescribe a $z$-dependent gain beyond $1$ m as $g(z)=1/[1+(z-z_1)/z_s]$, where $z_1=1$ m and $z_s=1.3$ m. The balance between gain, dispersion, and nonlinearity keeps the pulse peak power nearly constant during propagation, while the pulse energy exhibits algebraic growth and the pulse broadens temporally (Fig. 1c,d). The prescribed longitudinal gain produces an energy evolution given by $E(z)=E(z_1)[1+(z-z_1)/z_s]^{g_0z_s}$, with an energy-growth exponent $g_0z_s=1.3$ for the parameters considered here. The spectrogram in Fig. 1b illustrates the approximately linear chirp maintained during propagation as the pulse stretches in time while retaining nearly constant peak power. Over the propagation interval considered, the pulse energy is amplified by a factor of about 30. The output pulse exhibits a linearly chirped parabolic profile, making it suitable for subsequent compression. Importantly, the SPPS benefits from its nonlinear-attractor nature, meaning that its output characteristics are largely independent of the input pulse.

\begin{figure}[H]
    \centering
    \includegraphics[width=1\linewidth]{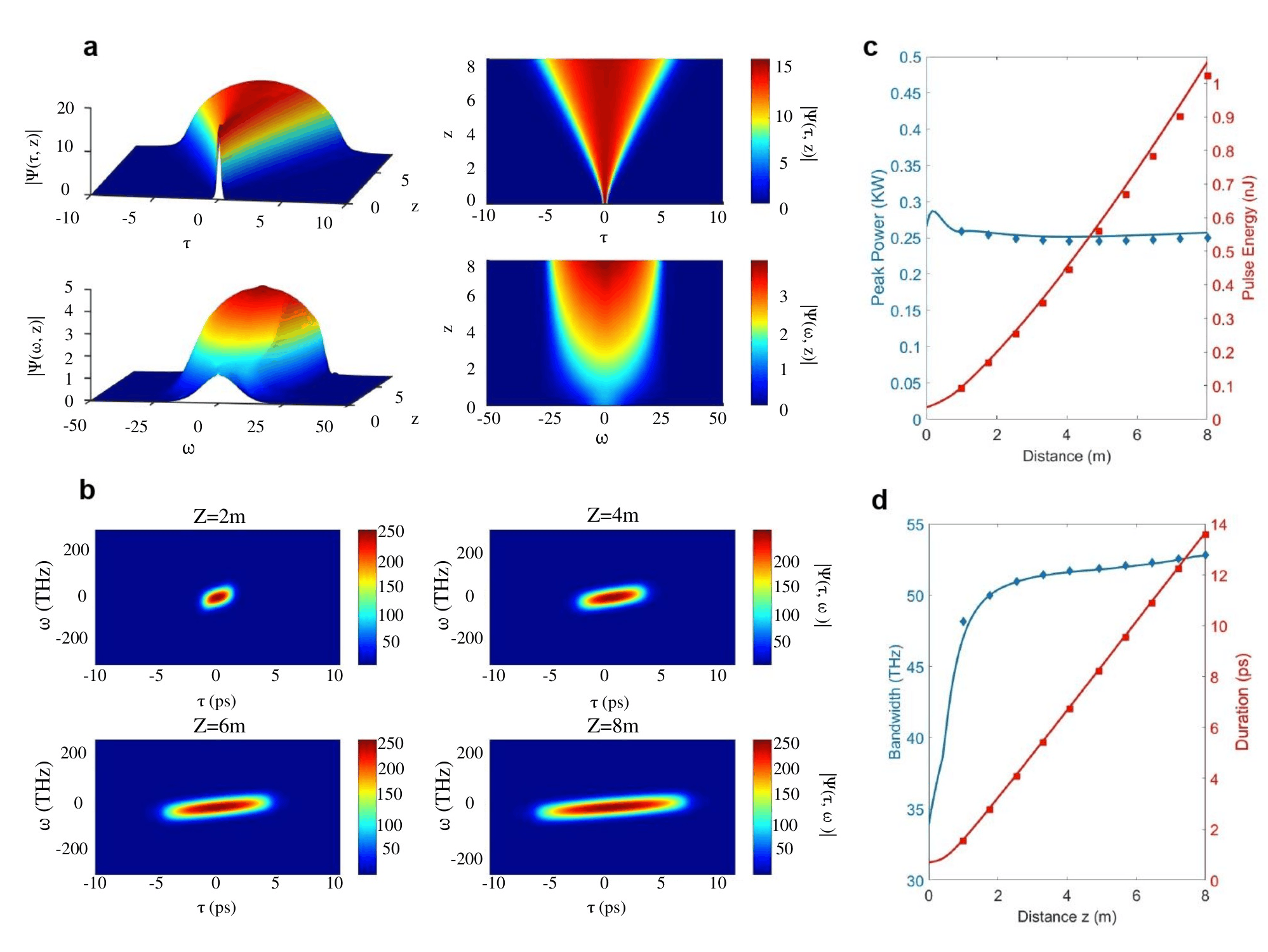}
    \caption{Stable peak power similariton (SPPS) dynamics in normal dispersion optical fibers. \textbf{(a)} SPPS generation in an 8-meter fiber amplifier using a 0.35 ps (FWHM) Gaussian input pulse with 35 pJ energy. Both temporal and spectral evolution is shown. \textbf{(b)} Spectrogram shows the approximately linear chirp characteristic of the similariton. \textbf{(c)} Balance between gain, dispersion, and nonlinearity maintains nearly constant pulse peak power during propagation, while the pulse energy exhibits algebraic growth. Blue rhombuses and red squares show results of the approximate semi-analytical model Eqs. (4,6). \textbf{(d)} Pulse duration increases due to temporal broadening, while the bandwidth saturates over propagation distance, exhibiting behavior distinct from traditional similaritons. Blue rhombuses and red squares represent our semi-analytical model Eqs. (4,6).}
    \label{fig:placeholder}
\end{figure}
We now turn to the frequency-limited gain, which becomes negative for shorter wavelength (see Fig.2a, solid blue line), as is typical for the GMNA regime \cite{sidorenko2019nonlinear}. We use the following model:
\begin{align}
\label{NLSE in canonical form}
g(\omega<\omega_0)=Ae^{-(\omega-\omega_0)^2/{2{\Delta\omega}^2}}\\
g(\omega>\omega_0)=A-B(\omega-\omega_0)^2
\end{align}
Here $A$ determines the peak gain and $B$ shapes the loss term, $\omega_0$ indicates the gain peak position, and $\Delta\omega$ defines the standard deviation of the Gaussian component. We input 0.5-ps-duration (FWHM) Gaussian pulse centered at $\lambda_0=1028$ nm, with a 35-pJ energy. The fiber parameters are $\gamma=3.91\times10^{-3}W^{-1}m^{-1}$ , $\beta_2=2\times10^{-3}ps^2m^{-1}$, corresponding to
a doped ytterbium fiber amplifier. The gain is constant between 0 m and 1 m, characterized by the parameters $A=1.75$m$^{-1}$, $B=0.01~\mathrm{m}^{-1}\mathrm{THz}^{-2}$, $\omega_0$ corresponds to 1028 nm, and $\Delta\omega=30THz$. At $z>z_0=1$ m the gain decreases with distance as $1/(1+(z-z_0)/z_s)$ with $z_0=1$m and $z_s=0.8$m. The fully numerical and the approximate
semi-analytical results are again in excellent agreement.

Initially, the pulse bandwidth is within the gain bandwidth (Fig.2a, orange and blue lines), enabling self-similar evolution.  During this stage, the pulse accumulates constant linear chirp. However, as the pulse bandwidth increases, the limited gain bandwidth impedes the formation of the stable peak-power similariton, leading to asymmetric temporal and spectral profiles (Fig. 2b, right column). The left column in Fig.2b shows the same simulation for the frequency-independent gain $g=1.75m^{-1}$ between 0m and 1m, which then decreases as $1/(1+(z-z_0)/z_s)$ beyond $z_0=1$m, with $z_s=0.8$m, where the SPPS emerges clearly. The comparison between the two columns of Fig. 2 establishes how finite spectral gain modifies the ideal SPPS solution. Importantly, this should not be interpreted as implying that all gain-managed nonlinear states evolve from the SPPS. Wang \cite{wang2026partial} have shown that realistic gain evolution can also produce gain-managed self-similar and gain-managed partial-parabolic regimes for different seed conditions. Our result is more specific: for the short-pulse conditions considered here, the frequency-independent SPPS provides a useful limiting solution from which the characteristic GMNA-like evolution emerges when the finite and asymmetric gain spectrum is introduced. In this sense, the SPPS picture and the regime classification of Ref[19] are complementary: the latter maps the different propagation regimes accessible as the seed and gain conditions are varied, whereas the present analysis isolates the longitudinal gain scaling associated with nearly stationary peak power.

\begin{figure}[H]
    \centering
    \includegraphics[width=1\linewidth]{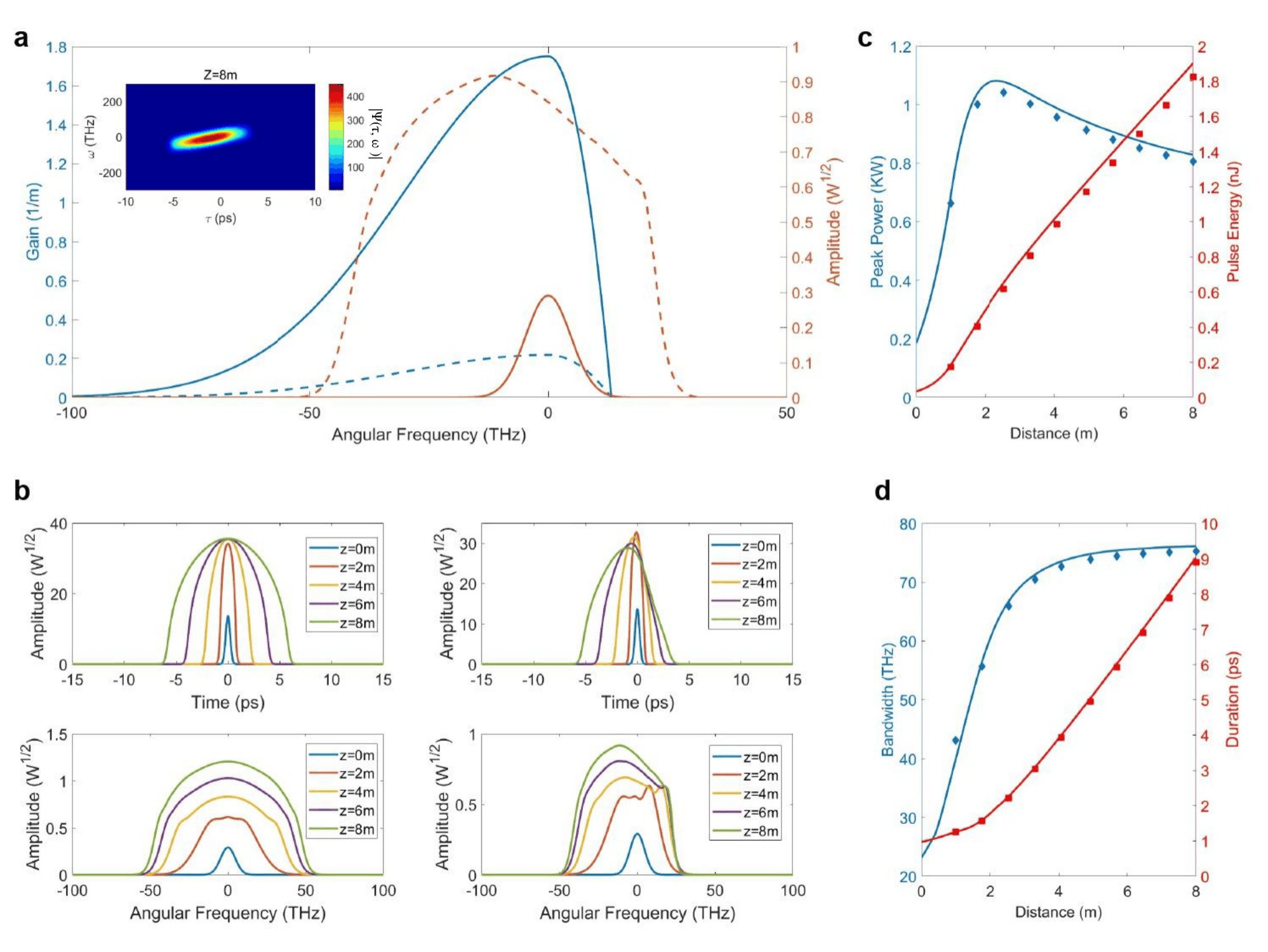}
    \caption{Stable peak power similariton transformation into the GMNA regime for frequency-dependent gain. Blue rhombuses and red squares show predictions of our semi-analytical model. \textbf{(a)} Solid orange and blue lines indicate the pulse (orange) and gain (blue) spectra at the input, while dashed lines show the same  at the output.\textbf{(b)} Stable peak power similariton (left) and the GMNA-type regimes (right) for a 0.5 ps (FWHM) input Gaussian pulse centered at 1028 nm with 35 pJ energy. The fiber parameters are $\gamma = 3.91 \times 10^{-3} \mathrm{W}^{-1} \mathrm{m}^{-1}$, $\beta_2 = 2 \times 10^{-3} \mathrm{ps}^2 \mathrm{m}^{-1}$. 
    %, and $T_R = 1.7\mathrm{fs}$
    For the similariton, the gain is $1.75 \mathrm{m}^{-1}$ from 0 m to 1 m and decreases as $1/(1 + (z - 1)/0.8)$ beyond 1m. For the frequency-dependent case, $A = 1.75 \mathrm{m}^{-1}$, $B = 0.01 ~\mathrm{m}^{-1}\mathrm{THz}^{-2}$, and $\Delta\omega = 30\mathrm{THz}$. \textbf{(c)} Peak power and pulse energy. \textbf{(d)} Pulse bandwidth and pulse duration. }
    \label{fig:placeholder}
\end{figure}
The finite gain bandwidth limits the pulse bandwidth at about 2-3 m, but the saturated pulse bandwidth of 75 THz well exceeds the gain bandwidth (see Fig.2a), dashed orange and dashed blue lines).  The pulse energy and duration continue to increase (Fig.2c,d), with the pulse maintaining nearly linear chirp. 

\begin{figure}[H]
    \centering
    \includegraphics[width=1\linewidth]{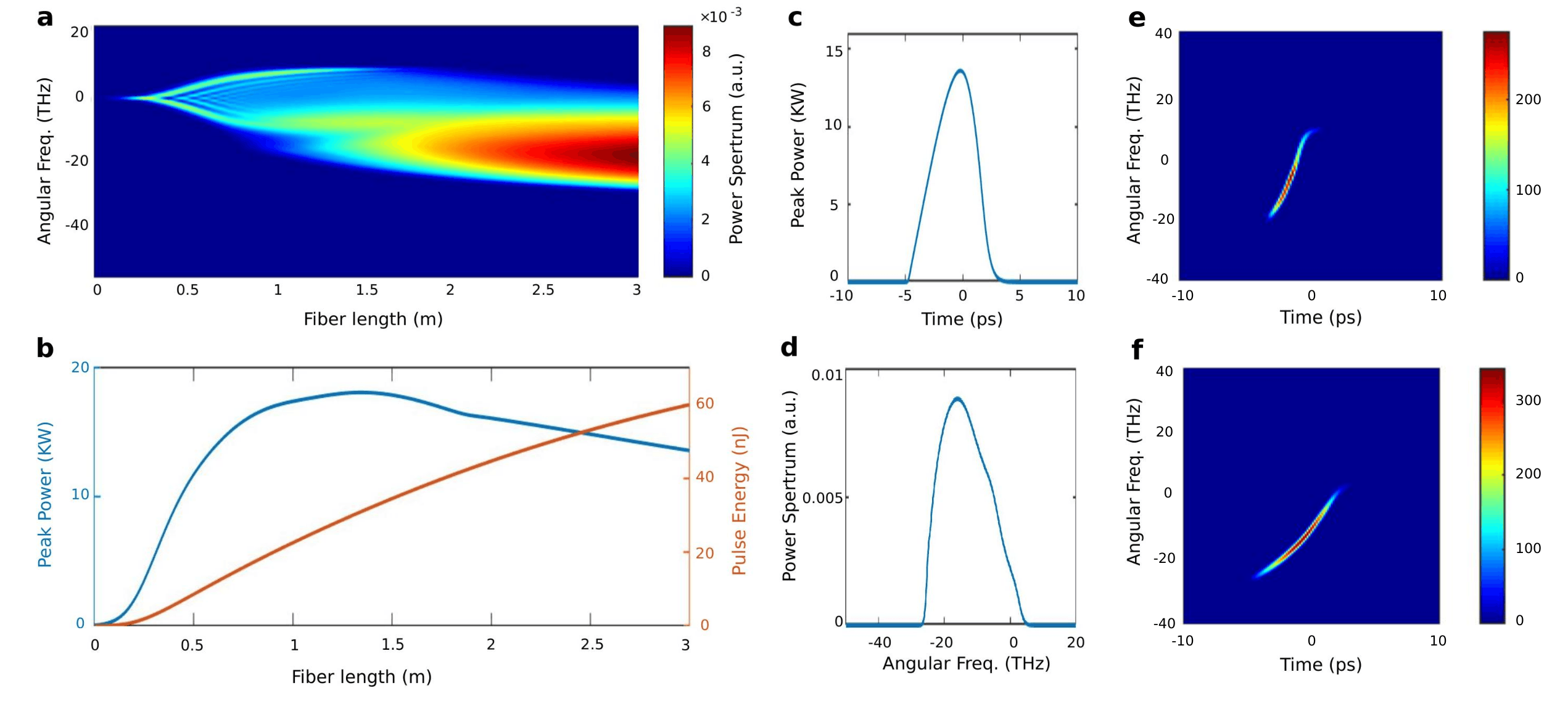}
    \caption{Pulse propagation in the GMNA regime in a double-clad Yb-doped fiber. (a) Spectral evolution of the pulse during amplification. (b) Evolution of the pulse peak power and energy along the gain fiber. (c) Temporal profile and (d) spectrum of the amplified pulse. (e) and (f) Spectrograms at 1.5 m and at the output of the gain fiber, respectively.}
    \label{fig:placeholder}
\end{figure}

The model clearly captures all the key features of the GMNA regime and allows its direct connection to the SPPS. Most importantly, we see that by carefully engineering the gain length dependence, one  can achieve an optimal balance 
between dispersion and nonlinearity to maximize the 
output pulse energy and spectral bandwidth. Last but not least, our semi-analytic model shows excellent agreement with the full simulation results (see Fig.1c and 1d), blue rhombuses and red squares).

Hybrid architectures in which a GMN stage provides a broadband seed for subsequent spectral shaping and amplification have also been demonstrated experimentally; for example, an all-fiber system recently produced 10.2-${\mu}J$ pulses compressed to 86 fs \cite{zhang2025generation}. Finally, we demonstrate a numerical example of a GMN amplifier that generates broadband pulses, followed by chirped-pulse amplification (CPA). Fig 3 presents simulations of a GMN amplifier consisting of 3 m of Coherent general-purpose Yb-doped double-clad fiber (PM-YDF-5/130-VIII), co-pumped at 976 nm. The simulations use the parameters $\gamma = 3.91 \times 10^{-3} \mathrm{W}^{-1} \mathrm{m}^{-1}$, $\beta_2 = 24 \times 10^{-3} \mathrm{ps}^2 \mathrm{m}^{-1}$ and $\beta_3 = 1 \times 10^{-4} \mathrm{ps}^3 \mathrm{m}^{-1}$. The numerical model includes second- and third-order dispersion, self-phase modulation, self-steepening, Raman scattering, and gain dynamics obtained by simultaneously solving the population inversion rate equations and the pulse propagation equations \cite{sidorenko2019nonlinear}.The coupled gain–propagation model belongs to the established class of self-consistent rate-equation/GNLSE models developed for ultrashort fiber amplifiers \cite{chen2023accurate,lindberg2016accurate}. The amplifier is seeded with a 500 fs transform-limited Gaussian pulse centered at 1030 nm with an energy of 0.03 nJ. Fig 3a shows the spectral evolution during amplification. Initially, pulse evolution is dominated by SPM. As amplification proceeds, the pulse evolves into the characteristic smooth, asymmetric profile of the GMN regime instead of undergoing wave breaking. The GMN regime is established when nonlinear spectral broadening balances gain shaping.
Fig 3b shows the evolution of the pulse peak power and energy along the gain fiber. Consistent with previous studies, the peak power exhibits a non-monotonic evolution arising from the dynamic interplay between gain and dispersion. The amplifier produces a 60 nJ output pulse, corresponding to an energy gain of 33 dB.
The temporal profile and optical spectrum of the amplified pulse are shown in Figs. 3c and 3d, respectively. Figs 3e and 3f present the pulse spectrograms at 1.5 m and at the output of the gain fiber, illustrating the gradual evolution toward the characteristic GMN pulse.
    Next, we investigate the amplification of the GMNA pulse in a CPA system. The primary objective of these simulations is to examine the trade-off between CPA gain and the gain-induced spectral narrowing associated with the Yb gain bandwidth. In the simulations, the GMNA pulse is first stretched using a group-delay dispersion (GDD) of $ 3 \times 10^{6} \mathrm{fs}^2 \mathrm{}$. The resulting stretched pulse, shown in Fig. 4a in time domain and spectral domain (Fig. 4b), has a full width at half maximum (FWHM) of 270 ps. 
% \begin{figure}[H]
%     \centering
%     \includegraphics[width=1\linewidth]{Fig4.png}
%     \caption{1.b: Maybe we could delete the word TL pulsepeak power and explain it in the text 2.c and d we could rearrange the position of these figures to make it looks more beautiful}
%     \label{fig:placeholder}
% \end{figure}

\begin{figure}[H]
    \centering
    \includegraphics[width=1\linewidth]{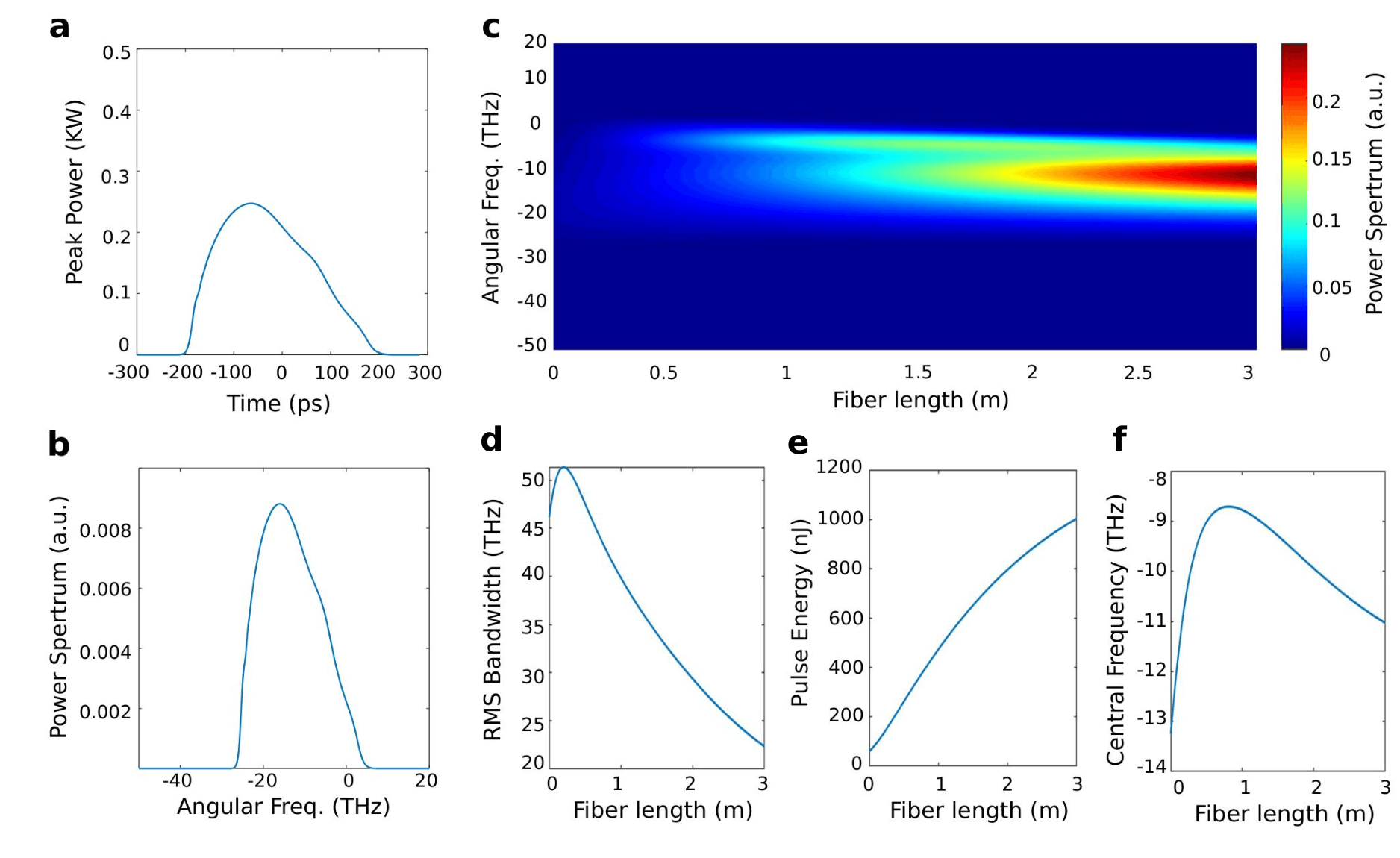}
    \caption{Demonstration of the GMNA-CPA system. (a) Temporal profile and (b) spectrum of the chirped pulse at the input of the gain fiber. (c) Spectral evolution during amplification. Evolution of the pulse (d) RMS spectral bandwidth, (e) pulse energy, and (f) center frequency along the gain fiber.}
    \label{fig:placeholder}
\end{figure}

Next, we use our numerical model to simulate amplification of the stretched pulse in the same Yb-doped fiber as in the previous section. Fig 4c shows the spectral evolution of the pulse during amplification. As expected, the spectral bandwidth gradually decreases due to the finite gain bandwidth of the Yb-doped fiber. The evolution of the pulse RMS spectral bandwidth is shown in Fig. 4d. The pulse enters the amplifier with an RMS bandwidth of 46 THz. During the initial stage of amplification (up to fiber length = 0.16 m), the bandwidth increases slightly because the pulse center frequency lies below the peak of the Yb gain spectrum. Beyond this point, the RMS bandwidth decreases monotonically as a result of gain narrowing. Since the Yb gain bandwidth is narrower than the pulse spectrum, frequencies near the gain peak experience higher amplification, leading to progressive spectral compression. Fig 4e shows the pulse energy as a function of propagation distance. Together, Figs. 4d and 4e illustrate the fundamental trade-off between pulse energy and spectral bandwidth in a GMNA-CPA system, providing a useful guideline for selecting the amplifier length according to the requirements of a particular application. For example, after 3 m of amplification, the pulse energy increases from 60 nJ to 1000 nJ, corresponding to a gain of 12.2 dB. At this point, the pulse retains an RMS bandwidth of 22 THz, corresponding to a transform-limited pulse duration of approximately 45 fs. The pulse center frequency also evolves during amplification due to gain reshaping. Figure 4f shows the center frequency as a function of propagation distance. Initially (up to approximately 0.9 m), the center frequency shifts toward higher frequencies because the pulse spectrum is pulled toward the gain maximum. For longer fiber lengths, the center frequency gradually shifts back to lower frequencies as gain narrowing progressively reduces the spectral wings. 

In conclusion, we have investigated the dynamics and generation of stable similaritons in optical fibers, providing both numerical and semi-analytical insights into their behavior under normal dispersion regimes. Our analysis  shows that the interplay of nonlinearity and dispersion in combination with gain management as a function of propagation length can lead to the formation of "stable peak power similaritons." 
Their ability to maintain stationary or even decreasing peak power during propagation, while increasing energy and exhibiting spectral broadening  beyond gain bandwidth present an exciting route for compact, high energy fiber-based  light sources. Recent work has shown that gain management can modify several distinct nonlinear amplification regimes, including self-similar and partial-parabolic amplification \cite{wang2026partial}. The present results provide a complementary limiting picture. Under spectrally flat but longitudinally managed gain, dispersion-driven pulse stretching can balance amplification to produce an SPPS with approximately stationary peak power and steadily increasing energy. When finite, asymmetric gain is introduced, this idealized solution evolves toward the characteristic behavior associated with GMNA. The SPPS can therefore be viewed as a useful reference state for understanding one important aspect of GMNA dynamics, rather than as a universal description of all gain-managed nonlinear amplification. 
We have also shown how the interplay of dispersion and self-phase-modulation  make similaritons a natural attractor during pulse propagation in gain media with normal dispersion. Finally, we numerically demonstrated a GMNA-CPA system and investigated the trade-off between CPA gain and gain-induced spectral narrowing of the GMNA pulse. Our simulations show that an all-single-mode-fiber GMNA-CPA system can deliver 1 $\mu$J pulses with a spectral bandwidth sufficient to support a transform-limited pulse duration of approximately 45 fs.  

\bibliography{bibliography}

\end{document}